\documentclass[twocolumn,showpacs,preprintnumbers,nofootinbib,prd,superscriptaddress,groupedaddress,10pt]{revtex4-1}
\usepackage{graphicx,amssymb,amsmath,amsthm,amsfonts,epsfig}
\usepackage[linktocpage]{hyperref}
\usepackage[usenames]{color}
\usepackage{epstopdf}

\usepackage{bm}
\usepackage{dcolumn}
\usepackage[latin1]{inputenc}
\usepackage{latexsym}
\usepackage{rotating}
\usepackage{hyperref}
\usepackage{color}
\usepackage{longtable}
\usepackage{enumerate}
\usepackage{mathtools}
\usepackage{url}
\renewcommand{\vec}[1]{\boldsymbol{#1}}

\newcommand{\ben}{\begin{enumerate}}
\newcommand{\een}{\end{enumerate}}

\def\be{\begin{equation}}
\def\ee{\end{equation}}
\def\bea{\begin{eqnarray}}
\def\eea{\end{eqnarray}}
\newcommand{\beq}{\begin{eqnarray}}
\newcommand{\eeq}{\end{eqnarray}} 
\newcommand{\ba}{\begin{align}}
\newcommand{\ea}{\end{align}}

\def\ba{\bar{a}}

\begin{document}

\title{Gravitational waves and higher dimensions:\\ Love numbers and Kaluza-Klein excitations}

\author{
Vitor Cardoso$^{1,2}$,
Leonardo Gualtieri$^{3,4}$,
Christopher J. Moore$^{5}$
}
\affiliation{${^1}$ CENTRA, Departamento de F\'{\i}sica, Instituto Superior T\'ecnico -- IST, Universidade de Lisboa -- UL,
Avenida Rovisco Pais 1, 1049 Lisboa, Portugal}
\affiliation{${^2}$ Institute for Theoretical Physics, University of Amsterdam, PO Box 94485, 1090GL, Amsterdam, The Netherlands
}
\affiliation{$^{3}$ Dipartimento di Fisica, ``Sapienza'' Universit\`a di Roma, Piazzale 
Aldo Moro 5, 00185, Roma, Italy}
\affiliation{$^4$ Sezione INFN Roma1, Piazzale Aldo Moro 5, 00185, Roma, Italy}
\affiliation{$^5$ School of Physics \& Astronomy and Institute for Gravitational Wave Astronomy, University of Birmingham, Birmingham, B15 2TT, UK}
\begin{abstract}
Gravitational-wave (GW) observations provide a wealth of information on the nature and properties of black holes.  Among
these, tidal Love numbers or the multipole moments
of the inspiralling and final objects are key to a number of constraints. Here, we consider these observations in the
context of higher-dimensional scenarios, with flat large extra dimensions.  We show that -- as might be anticipated, but
not always appreciated in the literature -- physically motivated set-ups are unconstrained by gravitational-wave
data. Dynamical processes that do not excite the Kaluza-Klein (KK) modes lead to a signal identical to that in
four-dimensional general relativity in vacuum .
In addition, any possible excitation of the KK modes is highly suppressed relative to the dominant quadrupolar term;
given existing constraints on the extra dimensions and the masses of the objects seen in gravitational-wave observations,
KK modes appear at post-Newtonian order $\sim 10^{11}$.
Finally, we re-compute the tidal Love numbers of spherical black holes in higher dimensions. We confirm that these are
different from zero, but comparing with previous computations we find a different magnitude and sign.
\end{abstract}

%\tableofcontents
%\end{widetext}
%\clearpage

\maketitle

%\tableofcontents
%%%%%%%%%%%%%%%%%%%%%%%%%%%%%%%%%%%%%%
\section{Introduction}
%%%%%%%%%%%%%%%%%%%%%%%%%%%%%%%%%%%%%%
The detection of gravitational waves (GWs) from compact binaries allows for new and sometimes unprecedented tests of
gravity in the strong-field, dynamical regime. These span a range of foundational issues, from constraining the speed
and mass of the graviton, to the nature of black hole
(BHs)~\cite{Yunes:2016jcc,Barack:2018yly,Cardoso:2019rvt,Baibhav:2019rsa,Cardoso:2019rou}.  The {\it dimensionality} of
spacetime is another fundamental question, first raised a century ago by the pioneering work of Kaluza, Klein and
  others.

Compactified extra dimensions are a prediction of the string theory/$M$-theory framework~\cite{Polchinski:1998rr}, but they are generally expected to have a compactification scale of the order of the Planck length $l_{\rm p}\sim10^{-33}$
cm, and thus not to be detectable by present observations and experiments. However, in the last two decades it has been
suggested that the existence of {\it large extra
  dimensions}\,\cite{ArkaniHamed:1998rs,Antoniadis:1998ig,Randall:1999ee}, with a compactification scale $L\gg l_{\rm
  p}$, could apparently solve some of the puzzles of the standard model of particle physics. In particular, the
longstanding hierarchy problem (i.e., the enormous gap between the standard model energy scale and the Planck energy
$E_{\rm p}\sim10^{19}$ GeV) could be addressed, since the effective Planck energy would be lowered to the TeV scale.
In these models gravity propagates in the $d$-dimensional {\it bulk}, while the standard-model fields are constrained to
a four-dimensional {\it brane} embedded in the bulk. Thus, gravity obeys the inverse-square law at
distances much larger than $L$, while it falls off as $\sim 1/r^{(d-2)/2}$ (see e.g.~\cite{Cardoso:2002pa}) at smaller
scales.

When the large-extra dimension scenario was first proposed, it was mostly unconstrained by observations and experiments:
laboratory tests of the inverse-square law only reached the millimiter scale, and particle colliders reached energies
well below the TeV. Now, such constraints have dramatically improved: lab-scale tests of gravity verified the
inverse-square law down to the micrometer scale~\cite{Sushkov:2011zz}, and the LHC has not found neither Kaluza-Klein
gravitons, not BHs, up to $\sim7.7$ TeV~\cite{Aad:2014wca,Sirunyan:2018ipj} (in processes with center-of-mass
energies up to $\sim13$ TeV). Even in this reduced parameter space, large extra dimensions are still an interesting and
well-motivated hypothesis, but - since larger energy scales are unreachable by present detectors - setting stronger
constraints is challenging.

With this in mind, it is natural to expect that current GW detectors have little to say about well-motivated
higher-dimensional universes: to test micrometer scale physics, one needs to probe frequencies of the order of $c/L~\sim
10^{14}\,{\rm Hz}$ or higher (see also~\cite{Andriot:2017oaz}). Such scales are unreachable by current GW detectors.

The purpose of the first part of this work, Section \ref{BS} below, is to quantify the expectation above. For that, we
will study perturbed BHs in higher dimensions, either by other BHs or simply in vacuum. We will show
that the only imprint of extra dimensions comes by way of fluctuations along the extra directions, which are nearly
impossible to be excited by astrophysical processes. Thus, the
characteristic modes of vibration or tidal Love numbers of astrophysical BHs remain the same as in
four-dimensional general relativity in vacuum. The second part of this work, Section~\ref{tln}, focuses on the calculation
of tidal Love numbers when the BH is much smaller than the scale of the extra dimensions. We show that these
numbers are nonzero, as previously shown~\cite{Kol:2011vg}, but we find a different magnitude and sign. 
This calculation shows that the vanishing of the Love numbers of BHs is
specific to four dimensions and not a general characteristic of the theory.

We use geometric units $c=G=1$, unless specified otherwise.
%%%%%%%%%%%%%%%%%%%%%%%%%%%%%%%%%%%%%%%%%%%%%%%%%%%%%%%%%%%%%%%%%%%%%%%%%%%%%
\section{Dynamics of black holes in higher dimensions\label{BS}}
%%%%%%%%%%%%%%%%%%%%%%%%%%%%%%%%%%%%%%%%%%%%%%%%%%%%%%%%%%%%%%%%%%%%%%%%%%%%%
In this article we focus on compactified spacetimes with {\it flat extra-dimensions}, such as those considered in the
Arkani Hamed-Dimopoulos-Dvali (ADD) model~\cite{ArkaniHamed:1998rs,Antoniadis:1998ig}; we shall not consider the cases
of warped extra-dimensions~\cite{Randall:1999ee} or of infinite volume extra dimensions~\cite{Randall:1999vf}.

In the ADD model, the spacetime is $d$-dimensional, with four non-compact and $d-4$ compactified dimensions. The
spacetime metric can thus be decomposed as
\begin{align}
  ds^2_d&=g_{AB}(y)dy^Ady^B=g_{\mu\nu}(x,z)dx^\mu dx^\nu\nonumber\\
  &+2g_{\mu i}(x,z)dx^\mu dz^i+g_{ij}(x,z)dz^idz^j\label{eq:metric}
\end{align}
were $\{y^A\}=\{x^\mu,z^i\}$, $x^\mu$ ($\mu=0,\dots3$) are the coordinates of the four-dimensional spacetime and $z^i$
($i=1,\dots,d-4$) are the coordinates of the extra dimensions. The standard model fields are constrained on a
four-dimensional brane ($\mathbf{z}=(z^1,\dots,z^{d-4})=\mathbf{0}$),
while gravity propagates in the $d$-dimensional bulk.

The compactification spacetime has the topology of a torus, therefore the coordinates $z^i$ are periodic, i.e. we
identify $z^i\to z^i+2\pi L$ (we assume for simplicity that the different compact directions have the same period).  It
is thus possible to expand the metric in a Fourier series
\begin{equation}
g_{AB}(x,z)=\sum_{m_z^1,\dots,m_z^{d-4}}g^{\vec m_z}_{AB}(x)e^{i\frac{\vec m_z\cdot\vec z}{L}}
\end{equation}
($\vec m_z=(m_z^1,\dots m_z^{d-4})$ integer numbers). The ``zero-mode'' ($\vec m_z=\vec0$) in this expansion,
$g_{AB}^{\vec 0}(x)$, is the product of a four-dimensional metric (solution of the four-dimensional Einstein's
equations) and of the metric of the $(d-4)$-torus:
\begin{equation}
ds^{2\,\vec0}_d=g^{\vec0}_{\mu\nu}(x)dx^\mu dx^\nu+\sum_{i=1}^{d-4}(dz^i)^2\,.\label{flat}
\end{equation}
The $\vec m_z\neq\vec 0$ terms, instead, are called {\it Kaluza-Klein (KK) modes} and are solution of {\it massive} field
equations, with masses of the order of $|\vec m_z|/L$.

Summarizing, in the flat extra dimensions scenario the spacetime is the sum of the zero-mode spacetime~\eqref{flat}, and
- if they are excited - of the KK modes, which depend on the extra dimensional coordinates.
%%%%%%%%%%%%%%%%%%%%%%%%%%%%%%%%%%%%%%%%%%%%%%%%%%%%%%%%%%%%%%%%%%%%%%%%%%%%%
\subsection{Black string and black brane solutions}\label{sec:homogeneous}
%%%%%%%%%%%%%%%%%%%%%%%%%%%%%%%%%%%%%%%%%%%%%%%%%%%%%%%%%%%%%%%%%%%%%%%%%%%%%
Different BH solutions are possible in this scenario~\cite{Kanti:2004nr,Cavaglia:2002si,Emparan:2008eg}.  If the BH is
much smaller than the scale $L$, it behaves approximately as living in noncompact dimensions, and can be well described
by the Tanngherlini solution, which is the $d$-dimensional extension of the Schwarzschild
solution~\cite{Tangherlini:1963bw,Myers:1986un}. However, we are interested in astrophysical objects, which are orders
of magnitude larger than the current upper bounds on $L$.

The simplest solution of {\it large BH} in the ADD scenario is the case in which the KK modes are not
excited:
\begin{equation}
ds^2_d=g^{BH}_{\mu\nu}(x)dx^\mu dx^\nu+\sum_{i=1}^{d-4}(dz^i)^2\,.\label{black_string}
\end{equation}
Here $g^{BH}_{\mu\nu}(x)dx^\mu dx^\nu$ is a BH solution of the four-dimensional Einstein's equations in vacuum. Indeed,
since the metric tensor~\eqref{black_string} does not depend on the coordinates $z^i$, the $D$-dimensional Einstein's
equations in vacuum, $R^{(D)}_{\mu\nu}=0$, reduce to the four-dimensional equations $R^{(4)}_{\mu\nu}=0$. Note that the
four-dimensional BH spacetime can be a stationary Schwarzschild or Kerr solution (in which case the solution is called a
{\it black string} if there is one compact dimension $[d=5]$, or a {\it black brane} if there are more compact
dimensions $[d>5]$) but it can also be a dynamical solution, such as the metric of a coalescing BH binary in vacuum.

Remarkably the black string/black brane solution, observed by a device constrained on the $\mathbf{z}=\mathbf{0}$ brane,
is in all respects indistinguishable from the corresponding BH solution of general relativity.
%%%%%%%%%%%%%%%%%%%%%%%%%%%%%%%%%%%%%%%%%%%%%%%%%%%%%%%%%%%%%%%%%%%%%%%%%%
\subsection{Fluctuations around a black string spacetime}\label{sec:fluc}
%%%%%%%%%%%%%%%%%%%%%%%%%%%%%%%%%%%%%%%%%%%%%%%%%%%%%%%%%%%%%%%%%%%%%%%%%%
To probe the extra dimensions, one needs to introduce non-homogeneities along the $z^i$ directions, i.e. to excite the
KK modes.  To this aim, let us consider a non-spinning black string, in which
$ds^2_4=-fdt^2+f^{-1}dr^2+r^2(d\vartheta^2+\sin^2\vartheta d\varphi^2)$ is the Schwarzschild spacetime ($f=1-2M/r$) and - for
simplicity - $d=5$, and perturb it. Since the extra dimension is periodic ($z\to z+2\pi L$), the fluctuation can be
decomposed as $\sim e^{ikz}$, where
\begin{equation}
k=\frac{m_z}{L}\,,\quad m_z=0\,,\pm 1\,,\pm2\dots
\end{equation}
Note that the only scenarios compatible with observations so far are those for which
\begin{equation}
L/M\ll 1\,,
\end{equation}
with $M$ the mass of an astrophysical BH. Fluctuations of such geometry were studied in Ref.~\cite{Kudoh:2006bp}. Let us
focus for example on the ``Regge-Wheeler'' (i.e., axial parity) sector, described by the master equation
\begin{equation}
f^2\Psi''+ff'\Psi'+(\omega^2-V)\Psi=0\,,
\end{equation}
with
\begin{align}
  V_{m_z}&=\frac{f}{r^2}\left(\frac{3f}{(1+{\cal R}^2)^2}+m_V(1+{\cal R}^2)\right.\nonumber\\
  &\left.+\frac{1-8f}{1+{\cal R}^2}+8f-2\right)\,.\label{RWstring}
\end{align}
Here %$f=1-2M/r$,
$m_V=l(l+1)-2$ and ${\cal R}^2=k^2r^2/m_V=m_z^2r^2/(L^2m_V)$.  Note that since $L/M\ll1$, outside the
horizon of an astrophysical BH ${\cal R}\gg 1$, unless $m_z=0$.

At low energies, when the fluctuations of the extra dimension
are not excited ($m_z=0$), the potential~\eqref{RWstring} reduces to
\begin{equation}
V_0=f\left(\frac{l(l+1)}{r^2}-\frac{6M}{r^3}\right)\,,
\end{equation}
which coincides with the Regge-Wheeler potential for axial-parity gravitational fluctuations of the standard
Schwarzschild geometry~\cite{Regge:1957td}.  This had to be the case, from the discussion in
Section~\ref{sec:homogeneous} above, and as had also been observed for other
models~\cite{Seahra:2004fg,Chakraborty:2017qve}.  This means that all the linearized dynamics are the same in such a
circumstance: BHs ring in the same way and are tidally deformed in the same way. No measurement is able to
distinguish between a four-dimensional BH and a $d-$dimensional black string when the fluctuations don't probe
the extra directions.

On the other hand, for $m_z\neq0$, one finds
\begin{equation}
V_{m_z}=f \,k^2\,,
\end{equation}
where we used the fact that for any interesting setup, ${\cal R}\gg 1$. In other words, the effective potential is
equivalent to that of a large-mass scalar around a Schwarzschild BH.  The solutions at large distance are
\begin{equation}
\Psi \sim e^{\pm i\sqrt{\omega^2-k^2}\,r}\,.
\end{equation}
Thus, for low-energy processes with $\omega<k$, the mode is simply not excited and no energy loss is observed far away
preventing any meaningful bound from GW observations.  It can easily be shown that, if these modes are excited, they
have frequency $\omega\gtrsim k$~\cite{Simone:1991wn,Cardoso:2005vk,Dolan:2007mj}. In other words, these modes show up
at frequencies (reinstating dimensionful units)
\begin{equation}
f\gtrsim \frac{m_zc}{2\pi L}>5\times 10^{13}\left(\frac{\mu {\rm m}}{L}\right)\,{\rm Hz}\,.
\end{equation}
In the case of more than one extra dimensions, i.e. $d>5$, the results are qualitatively similar.

%%%%%%%%%%%%%%%%%%%%%%%%%%%%%%%%%%%%%%%%%%%%%%%%%%%%%%%%%%%%%%%%%
\subsection{Excitation of KK modes and GW astronomy}
%%%%%%%%%%%%%%%%%%%%%%%%%%%%%%%%%%%%%%%%%%%%%%%%%%%%%%%%%%%%%%%%%
The relevant question is then if such massive modes are excited during astrophysical processes\,\footnote{We do not
  consider particle physics processes. Indeed, we know that KK gravitons are not excited in processes with energies
  below $\sim13$ TeV~\cite{Sirunyan:2018ipj}; to our knowledge, we do not expect processes with this - or higher - energy
  scale to occur in astrophysical phenomena involving stars or BHs.}.
A simple prototypical
model shows that such modes are extremely suppressed. Instead of the full gravitational problem, let us consider the
dynamics describing the interaction between a black string and a scalar charge, which is assumed to be a small
perturbation in the black string spacetime, and which excites a scalar field $\Phi$ to which it couples.
This problem is described by
\begin{equation}
\nabla_A\nabla^A\Phi=\alpha {\cal T}\,,
\end{equation}
with ${\cal T}$ the trace of the stress-energy tensor of the charge and $\alpha$ some coupling constant.  If ${\cal T}$
is homogeneous in $z$, then the only mode excited is $m_z=0$. Indeed,
Fourier-expanding $\Phi$ and ${\cal T}$ in $z$, the $m$-th mode of the source sources the corresponding mode of the field.

Suppose instead that the source is maximally inhomogeneous, i.e., a pointlike source moving along a geodesic. For head-on collisions ${\cal T}=-\frac{m_p}{\sqrt{-g}U^t}\delta(r-R(T))
\delta(\cos\vartheta-1)\delta(\varphi)\delta(z)$, while for a particle on a circular motion at $r=r_0$,
${\cal  T}=-\frac{m_p}{\sqrt{-g}U^t}\delta(r-r_0)\delta(\cos\vartheta)\delta(\varphi-\Omega t)\delta(z)$.  Projecting the field
into spherical harmonics and Fourier-decomposing,
\begin{align}
  \Phi(t,r,\vartheta,\varphi,z)&=\frac{1}{\sqrt{2\pi}}\int d\omega e^{i\omega t}\times\nonumber\\
  &\times\sum_{l,m,m_z}\frac{\Phi_{lmm_z}(r)}{r}Y_{lm}(\vartheta,\varphi) e^{i\frac{m_zz}{L}}
\end{align}
yields the following ordinary differential equation
\begin{equation}
f^2\Phi_{lm\,m_z}''+ff'\Phi_{lm\,m_z}'+\left(\omega^2-V\right)\Phi_{lm\,m_z}=m_p\frac{f}{r\,LU^t}{\cal S}\,,
\label{eqpert}
\end{equation}
where $U^t=dt/d\tau$,
\begin{equation}
V=f\left(\frac{l(l+1)}{r^2}+\frac{f'}{r}+k^2\right)
\end{equation}
(we remind that $k=m_z/L$), and the source term is
\begin{equation}
  {\cal S}= \frac{Y_{l0}(0)e^{i\omega T(r)}}{\sqrt{2\pi}\, dR/dt}
\end{equation}
in the head-on case, or
\begin{equation}
  {\cal S}=\sqrt{2\pi}\,Y_{lm}(\pi/2)\delta(r-r_0)\delta(\omega-m\Omega)
\end{equation}
in the circular case. 

Head-on collisions only excite modes with $M\omega\sim 1$, and the energy output dies exponentially at large
frequencies, $dE/d\omega\sim e^{-M\omega}$~\cite{Zerilli:1971wd}. Since we need $\omega>k$ to have propagating modes
(see Sec.\ref{sec:fluc}), the high-frequency radiation is exponentially suppressed for head-on collisions, i.e. 
$E\sim e^{-M/L}$.

In the case of circular motion, which in our setup would describe well the physics of an extreme-mass-ratio inspiral,
%we see that
the excited states only contribute at $m\Omega \gtrsim k$.
Using the fact that $\Omega=(M/r_0^3)^{1/2}$ in circular motion and that the highest angular velocity is at the
innermost stable circular orbit, we find that the azimuthal index $m$ must satisfy
\begin{equation}
m\gtrsim 2.2\times 10^{10}\frac{M}{M_{\odot}}\frac{\mu{\rm m}}{L}\,.\label{eq:m_crit}
\end{equation}
The compact extra dimension introduces a new fundamental lengthscale $L$ breaking the scale invariance of general relativity. 
Therefore, the post-Newtonian order at which corrections from the extra dimension enter the signal depends on the mass of the system.
The contribution from the dominant $l=m$ modes is suppressed by $v^{2(m-2)}$ relative to the dominant quadrupole term~\cite{Poisson:1993vp}.
Given astrophysically relevant masses and the existing constraints $L\lesssim \mu m$ we see that the corrections only enter at extremely high orders of
\begin{equation}
v^{N}\,,\quad\textrm{where} \quad N\gtrsim 4.4\times 10^{10} \frac{M}{M_{\odot}}\frac{\mu{\rm m}}{L}\,.
\end{equation}
This computation readily extends to the case of multiple extra dimensions, with qualitatively similar results.

Although this is a toy model for a real astrophysical process, taken with the previous example it provides strong
evidence that GW astronomy cannot be used to impose meaningful constraints on extra-dimensional scenarios with flat
extra directions.

%%%%%%%%%%%%%%%%%%%%%%%%%%%%%%%%%%%%%%%%%%%%%%%%%%%%%%%%%%%%%%%%%
\subsection{Perturbations around a flat background}
%%%%%%%%%%%%%%%%%%%%%%%%%%%%%%%%%%%%%%%%%%%%%%%%%%%%%%%%%%%%%%%%%
Similar conclusions apply to considerations of weak gravitational fields.  The $d$-dimensional Einstein field equations
can be written in trace-reversed form as
\begin{equation}
    R_{AB} = 8\pi\left(T_{AB}-\frac{1}{d-2}g_{AB}T\right) \,,
\end{equation}
where $R_{AB}$ is the Ricci tensor derived from the full metric $g_{AB}(y)$ (see Eq.~\eqref{eq:metric}).  For weak
gravitational fields, the metric can be expanded about a flat Minkowski background, $g_{AB}(y) = \eta_{AB} + h_{AB}(y)$,
yielding the following (harmonic gauge) linearised field equations for the metric perturbation:
\begin{equation} \label{eq:lin_field_eqns}
    \partial_{C}\partial^{C} h_{AB} = -16\pi\left(T_{AB}-\frac{1}{d-2}g_{AB}T\right) \,.
\end{equation}
Let us first consider the vacuum field equations, $T_{AB}=0$.  Here it is additionally possible to enforce
\emph{transverse-traceless} gauge conditions. There exist GW solutions to these equation propagating along one of the
infinite directions with $d(d-3)/2$ \cite{Cardoso:2002pa} distinct GW polarization states; when $d=5$, in addition to
the familiar $+$ and $\times$ states on the brane there exist 3 additional states with components along the compact bulk
direction. 

Let us now consider the perturbation equations with source, Eq.~\eqref{eq:lin_field_eqns}. They may be solved using the
Green function method. Following \cite{Barvinsky_Solodukhin_echo_extra_dim}, the $d=5$ dimensional Green's function can
be written as a tower of KK modes,
\begin{align} \label{eq:KKtower}
    G^{(5)}(y,y') = \frac{1}{L}\sum_{n=-\infty}^{\infty} G^{(4)}_{m_n}(x,x')\textrm{e}^{-2\pi\textrm{i} n (z-z')/L} \,,
\end{align}
where the KK modes have masses $m_n=2\pi n/L$ and $G^{(4)}_{m}(x,x')$ is the four-dimensional Green's function for the
wave equation with mass $m$. The effect of the excitations of the $n^\textrm{th}$ KK mode is localized to within a
distance $\sim L/n$ of the source. Equivalently, the KK modes are excited only at frequencies $\omega\gtrsim n/L$; at
larger distances (or smaller frequencies) the effects are exponentially suppressed. Thus, one recover the result
\eqref{eq:m_crit} also for gravitational waves. Therefore at astrophysical relevant distances the solution is governed
by the massless term $G^{(4)}(x,x')$ (with no $z$ dependence) and the four-dimensional solution is recovered
identically. In the case of multiple extra dimensions, i.e. $d>5$, the results are again qualitatively
similar. Therefore, we do not expect GW observations to be a good probe for flat extra dimensions, whether the GWs are
sourced by vacuum BHs or by matter.

%%%%%%%%%%%%%%%%%%%%%%%%%%%%%%%%%%%%%%%%%%%%%%%%%%%%%%%
\section{Tidal Love numbers of Tangherlini black holes\label{tln}}
%%%%%%%%%%%%%%%%%%%%%%%%%%%%%%%%%%%%%%%%%%%%%%%%%%%%%%%
The previous Section dealt with geometries that could be astrophysically relevant. We now deal with geometries for which
the BH is much smaller than the size of the extra dimensions. In this case the static BH is symmetric under rotations on
the $(d-2)$-sphere, and is described by the Tangherlini solution~\cite{Tangherlini:1963bw}. Such BHs, as far as we know,
can not be astrophysical objects but this study serves a purpose. Einstein equations have a nontrivial content in higher
dimensions and thus BHs may have different properties at different values of $d$. A particularly intriguing property of
four-dimensional vacuum BH spacetimes is that their tidal Love numbers are zero~\cite{1983grr..proc...58D,Binnington:2009bb,Damour:2009vw,Porto:2016pyg,Porto:2016zng,Rothstein:simons,Cardoso:2019rvt}. We
wish to understand if such property is intrinsically four-dimensional, confirming or disproving earlier
results~\cite{Kol:2011vg}.
%Moreover, understanding tidal deformations of higher dimensional BHs is useful to address -
%through a dimensional regularization procedure - conceptual issues arising in the definition of tidal deformability in
%four dimensions (see e.g.~\cite{Damour:2012yf,Pani:2015hfa}).

The tidal deformability of Tangherlini BHs has been computed in Ref.~\cite{Kol:2011vg} (hereafter KS). Our results are
qualitatively in agreement with those of KS: - we also find non-vanishing tidal deformability when $2l/(d-3)$ is not integer -
but we find different numerical values:
\begin{eqnarray}
  \lambda&=&-\frac{d-2+l}{l-1}\lambda^{KS}  \nonumber\\
 &=&\frac{d-2+l}{l-1} \frac{\Gamma(\hat l)\Gamma(\hat l+2)}{\Gamma(\hat l+\frac{1}{2})\Gamma(\hat l+\frac{3}{2})}
  \tan\pi\hat l\left({r_s^{\hat d}}/{4}\right)^{2\hat l+1}\label{lovenumber}
\end{eqnarray}
where $\lambda$ is the tidal deformability, $\lambda^{KS}$ is the value computed in KS, $d$ is the spacetime dimension,
$l$ is the harmonic index, $\hat d=d-3$, $\hat l=l/\hat d$, and $r_s$ is the Schwarzshild radius. Strictly speaking, the Love numbers are dimensionless quantities obtained by a suitable rescaling of
the tidal deformabilty, but for simplicity of notation we simply call ``Love numbers'' the tidal deformability itself.

The Tangherlini spacetime is given by (we follow the formalism and the notation of Kodama and
Ishibashi~\cite{Kodama:2003kk}, hereafter KI)
\begin{equation}
ds^2=-f(r)dt^2+f^{-1}(r)dr^2+r^2d\sigma^2_n=g^{(0)}_{\mu\nu}dx^\mu dx^\nu
\end{equation}
where we have defined $n=d-2$ (i.e., $n=\hat d+1$),
\begin{equation}
f(r)=1-\left(\frac{r_s}{r}\right)^{n-1}=1-\frac{2M}{r^{n-1}}
\end{equation}
and $d\sigma^2_n$ is the metric of the $n$-sphere. We split the coordinates as $x^\mu=(z^a,y^i)$ where $a=0,1$,
$z^a=t,r$ and $y^i$ ($i=1,\dots n$) are the coordinates on the $n$-sphere, whose metric is
$d\sigma_n^2=\gamma_{ij}dy^idy^j$. We also denote with $D_a$ the covariant derivative on the two-dimensional manifold
$(t,r)$.

The perturbed metric is
\begin{equation}
  g_{\mu\nu}=g^{(0)}_{\mu\nu}+\delta g_{\mu\nu}\,.
\end{equation}  
The metric perturbations belong to three classes: scalar-like, vector-like and tensor-like.  Following Geroch and
Hansen~\cite{Geroch:1970cd,Hansen:1974zz}, we expect that the multipole moments of a static spacetime are defined in
terms of the norm of the timelike Killing vector, i.e. $\sqrt{-g_{00}}$. Since vector-like and tensor-like perturbations
of $g_{00}$ identically vanish, we only consider scalar-like perturbations, which are decomposed as:
\begin{equation}
  \delta g_{ab}=f_{ab}{\cal S}\,,~\delta g_{ai}=rf_a{\cal S}_i\,,~\delta g_{ij}=2r^2(H_L\gamma_{ij}{\cal S}+H_T{\cal S}_{ij})\,,
\end{equation}
where ${\cal S}$, ${\cal S}_i$ and ${\cal S}_{ij}$ are scalar, vector and tensor spherical harmonics on the $n$-sphere,
and the perturbation functions $f_{ab}$, $f_a$, $H_L$, $H_T$ are in general functions of $(t,r)$. Since we consider {\it
  static perturbations}, we assume they are only functions of $r$.

Following KI, we introduce the gauge-invariant perturbations
\begin{align}
  F&=H_L+\frac{1}{n}H_T+\frac{1}{r}D^arX_a\nonumber\\
  F_{ab}&=f_{ab}+D_aX_b+D_bX_a\label{gaugeinvpert}
\end{align}
where 
\begin{equation}
X_a=\frac{r}{\kappa}\left(f_a+\frac{r}{\kappa}D_aH_T\right)
\end{equation}
and
\begin{equation}
  \kappa^2=l(l+n-1)\,,~~~l=0,1,\dots
\end{equation}
are the eigenvalues of the scalar harmonics. Note that it is alway possible to choose a ``generalized Regge-Wheeler
gauge'' in which $f_a=H_T=0$; in this gauge, $F=H_L$ and $F_{ab}=f_{ab}$. The four quantities defined in
Eq.~\eqref{gaugeinvpert} are dynamically constrained by one of Einstein's equations, which yields $ 2(n-2)F+F^a_a=0$;
therefore, there are three independent perturbation functions, which are solution of a first-order system of partial
differential equations (see KI for details). In the static case, one of these equations yields $F_{tr}=0$ (modulo a
constant which can be set to zero with a gauge choice), and thus the independent quantities are reduced to two, which we
call (extending the standard notation of the four-dimensional case)
\begin{equation}
H_0=F^t_{~t}\,,~~~~H_2=F^r_{~r}\,.
\end{equation}

As discussed in KS, the Love numbers can be obtained in terms of a scalar field $\phi(r,\vartheta,\varphi)$, which is a
suitably defined combination of the metric perturbations. Each harmonic component $\phi_l(r)$ of the scalar field is
solution of a second-order differential equation; the general solution is a combination of two independent solutions,
which, expanded in powers of $r_s/r$, have the form:
\begin{align}
  \phi_l^{(1)}&=r^l\left(1+c_{11}\left(\frac{r_s}{r}\right)^{\hat d}+
  c_{12}\left(\frac{r_s}{r}\right)^{2\hat d}+\dots\right)\label{phi1}\\
  \phi_l^{(2)}&=\frac{1}{r^{l+\hat d}}\left(1+c_{21}\left(\frac{r_s}{r}\right)^{\hat d}+
  c_{22}\left(\frac{r_s}{r}\right)^{2\hat d}+\dots\right)\,,\label{phi2}
\end{align}
with $c_{1i}$, $c_{2i}$, ($i=1,\dots$) constants depending on $\hat d$ and $l$.  Here $\phi_l^{(1)}$ describes a test
tidal field, while $\phi_l^{(2)}$ describes the multipole moment response of the BH to the test field. The general
solution is a combination
\begin{equation}
  \phi_l(r)=const.(\phi^{(1)}_l(r)+\lambda_l\phi^{(2)}_l(r))\,,\label{deflambda}
\end{equation}
and the constant $\lambda_l$ is the $l$-th Love number. 

The Love numbers determined with this approach, although qualitatively in agreement with those found by KS, have different numerical values (see Eq.~\eqref{lovenumber}).  The difference between our derivation and that of KS is in the choice of the scalar field
$\phi$.

%As discussed by Geroch and Hansen~\cite{Geroch:1970cd,Hansen:1974zz}, the multipole moments of static BHs can be
%defined, in a gauge-invariant way, from the asymptotic expansion of the norm of the (asymptotically) timelike Killing
%vector

The gauge-invariant multipole moments of static BHs can be defined in terms of the asymptotic expansion of the norm of
the (asymptotically) timelike Killing vector $\xi^\mu$~\cite{Geroch:1970cd}
\begin{equation}
\phi=(-\xi^\mu\xi_\mu)^{1/2}-1=\left(f\left(1-\sum_{lm}H_{0\,lm}Y^{lm}\right)\right)^{1/2}-1
\end{equation}
thus, at linear order in the perturbations, the harmonic components with $l\ge2$ are
\begin{equation}
\phi_l=-\frac{1}{2}H_{0\,l}
\end{equation}
(we leave implicit the index $m$ since it does not affect the perturbation equations). Therefore, we combined the KI equations
into a single second-order differential equation in the perturbation function $H_l(r)\equiv H_{0\,l}$ (Eq.~\eqref{eq:H} in
Appendix~\ref{app:equation}). In the four-dimensional case ($n=2$), Eq.~\eqref{eq:H} reduces to 
\begin{equation}
H''_l+\frac{2r-r_s}{r(r-r_s)}H'_l-\frac{k^2r(r-r_s)-r_s^2}{r^2(r-r_s)^2}H=0\label{eq:hind}
\end{equation}
which, expressed in the variable in $2r/r_s-1$, coincides with the Hinderer equation (Eq.~(18) of~\cite{Hinderer:2007mb}).

Eq.~\eqref{eq:H} does not have an obvious solution in terms of special functions, therefore we have solved it numerically,
for each choice of $(n,l)$. We found that in the near-horizon limit the solution has the form
\begin{equation}
  H(r)=C_0+\frac{C_1}{r-r_s}+O(r-r_s)\,.
\end{equation}
Setting $C_1=0$ (i.e., imposing regularity at the horizon) and numerically integrating up to $r\gg r_s$, we found that
the solution (modulo normalization) has the form
\begin{equation}
H=H_l^{(1)}+\lambda H_l^{(2)}\label{H12}
\end{equation}
where
\begin{align}
  H_l^{(1)}&=r^l\left(1+c_{11}\left(\frac{r_s}{r}\right)^{\hat d}+
  c_{12}\left(\frac{r_s}{r}\right)^{2\hat d}+\dots\right)\label{H1}\\
  H_l^{(2)}&=\frac{1}{r^{l+\hat d}}
  \left(1+c_{21}\left(\frac{r_s}{r}\right)^{\hat d}+
    c_{22}\left(\frac{r_s}{r}\right)^{2\hat d}
  +\dots\right)\label{H2}
\end{align}
and $c_{1i}$, $c_{2i}$ are constants depeding on $n,l$.  By comparing our numerical solution with Eq.~\eqref{H12} we can
compute the Love number $\lambda$ for each choice of $l,n$. Our results are well described (withing $\lesssim1\%$, due
to numerical truncation error) by the analytical expression shown in Eq.~\eqref{lovenumber}.

The choice of the scalar field in KS is different, i.e. (leaving implicit the harmonic index $l$)
\begin{equation}
  \hat Y= r^{n-1}(F^r_{~r}-2F)\,.
\end{equation}
As shown in~\cite{Kodama:2003jz}, in the static limit the equation for this perturbation has a very simple form:
\begin{equation}
  f\frac{d^2\hat Y}{dx^2}-2\frac{d\hat Y}{dx}-\frac{\hat l({\hat l}+1)}{x^2}\hat Y=0\,,\label{eq:tY}
\end{equation}
where $x=(r_s/r)^{n-1}$.  This equation was analytically solved in KS in terms of special functions, finding an explicit
expression of the Love number $\lambda^{KS}$. The claim that $\lambda^{KS}$ is the Love number of the BH is bases on the
remark, in KS, that since there is only one gauge-invariant scalar the limit of $\hat Y$ and $\phi$ must be proportional
to each other. However, we think this is not the case. Indeed, the equation for $\hat Y$ does not reduce to the Hinderer
equation~\eqref{eq:hind}. This explains the discrepancy between our results and those of KS,
i.e. $\lambda=-(n+l)/(l-1)\lambda^{KS}$. Still, our computation confirms the qualitative results of KS, i.e. that
$\lambda=0$ for integer $\hat l$. In four dimensions $\hat l$ is integer for all values of $l$, and thus the Love
numbers vanish. This confirms the current understanding of four dimensional, spherically symmetric black holes (see
e.g.~\cite{Binnington:2009bb,Damour:2009vw,Pani:2015hfa,Gurlebeck:2015xpa}). In five dimensions, instead, the Love
numbers vanish only for even values of $l$.

We also find that $\lambda$ is finite and nonvanishing otherwise, except for half-integer $\hat l$, where
Eq.~\eqref{lovenumber} breaks down. It is easy to show that new logarithmic terms appear and that Love numbers are
ill-defined (in the way we defined them above, since the logarithmic terms are dominant). This aspect was also discussed
in KS. Naturally, Eq.~\eqref{lovenumber} cannot be taken at face value at these points, as it would yield a diverging
Love number: this would indicate a non-vanishing hair at linear level, which can be proven not to exist for such
background solutions. In fact, a numerical solution of the differential equation immediately shows that nothing blows up
at half-integer values of $\hat l$.

%%%%%%%%%%%%%%%%%%%%%%%%%%%%%%%%%%%%%%%%%%%%%%%%%%%%%%%%%%%%%%%%%%%%%%%%%%%%%%
\section{Discussion}
%%%%%%%%%%%%%%%%%%%%%%%%%%%%%%%%%%%%%%%%%%%%%%%%%%%%%%%%%%%%%%%%%%%%%%%%%%%%%%

The underlying dimensionality of the spacetime is a key issue. There are simple tests that can be done using GW
observations; for example, simple energy-conservation arguments imply that in a truly $d-$dimensional universe the
amplitude of GWs does not fall-off as $h\sim 1/r$ but rather as $\sim 1/r^{(d-2)/2}$~\cite{Cardoso:2002pa}. Since the
GW170817 standard siren measurement of the Hubble constant is consistent with expectations~\cite{Abbott:2017xzu}, one
has a strong constraint on the dimensionality $d$~\cite{Deffayet:2007kf,Pardo:2018ipy}. However appealing, these are
naive attempts in light of already-existing bounds, and most of them stumble upon a very simple fact: gravity obeys the
inverse square law to a very good precision, on scales ranging from a micrometer to galactic scales.  Notwithstanding,
the {\it apparent} four-dimensional nature of our universe is naturally explained away when the extra dimensions have a
typical size $L\lesssim \mu {\rm m}$~\cite{ArkaniHamed:1998rs,Antoniadis:1998ig} or are warped within such a
scale~\cite{Randall:1999vf,Randall:1999ee}. In such scenarios, the inverse square law is recovered at distances much
larger than the compactification radius.

We have shown in Section \ref{tln} that such scenario would have, as a matter of principle,
imprints on the BH physics: small BHs have non-vanishing tidal Love numbers, unlike their four-dimension kins. This
confirms earlier calculations~\cite{Kol:2011vg}, while correcting their precise magnitude and sign.
However, in Section~\ref{BS}
we have shown that GW observations have little to say about realistic higher-dimensional scenarios with flat extra directions.

Our construction says nothing about warped scenarios. There have been some claims in the literature that GW observations
can impose strong constraints on such setups~\cite{McWilliams:2009ym}. Such claims were based on early suspicions that
there were no stationary black hole solutions in these setups~\cite{Emparan:2002px}, which supposedly led to a large
Hawking evaporation. However, stationary black holes solutions have since been
built~\cite{Figueras:2011gd,Abdolrahimi:2012qi} and thus the validity of these constraints is at least
questionable.

\begin{acknowledgments}
We thank Barak Kol and Rafael Porto for useful discussions.
V. C. is partially supported by the Van der Waals Professorial Chair at the University of Amsterdam.
V.C.\ acknowledges financial support provided under the European Union's H2020 ERC 
Consolidator Grant ``Matter and strong-field gravity: New frontiers in Einstein's 
theory'' grant agreement no. MaGRaTh--646597.
This project has received funding from the European Union's Horizon 2020 research and innovation programme under the Marie Sklodowska-Curie grant agreement No 690904.
We acknowledge financial support provided by FCT/Portugal through grant PTDC/MAT-APL/30043/2017.
The authors would like to acknowledge networking support by the GWverse COST Action 
CA16104, ``Black holes, gravitational waves and fundamental physics.''
\end{acknowledgments}

\appendix
%%%%%%%%%%%%%%%%%%%%%%%%%%%%%%%%%%%%%%%%%%%%%%%%%%%%%%%%%%%%%%%%%%%%%%%%%%%%%%
\section{Equation for static perturbations of Tangherlini black holes}\label{app:equation}
%%%%%%%%%%%%%%%%%%%%%%%%%%%%%%%%%%%%%%%%%%%%%%%%%%%%%%%%%%%%%%%%%%%%%%%%%%%%%%
The equation for the perturbation function $H_l(r)=H_{0\,l}$ of Tangherlini black holes discussed in Section~\ref{tln},
in the case of static perturbations, is
\begin{equation}
H''_l(r)+c_1H'_l(r)+c_0H_l(r)=0\label{eq:H}
\end{equation}
where
\begin{widetext}
\begin{align}
  c_1&=\left[2 f^2 (n-1) n \left(r^n \left(n^2 ( {{f'}} r-2)+n \left(-2  {{f'}} r+2 \kappa ^2+2\right)-2 \kappa
    ^2\right)-2 M n \left(n^2-3 n+2\right) r\right)\right.\nonumber\\
&\left.    +f  {{f'}} r^{n+1} \left(n^3 (3  {{f'}} r-4)+n^2
   \left(-8  {{f'}} r+4 \kappa ^2+8\right)+4 n \left( {{f'}} r-2 \kappa ^2-1\right)+4 \kappa ^2\right)+2
        {{f'}}^3 (n-2) n r^{n+3}\right]\nonumber\\
&\times  \left[f r \left(2 f (n-1) \left(r^n \left(n^2 ( {{f'}} r-2)+n \left(-2  {{f'}}
   r+2 \kappa ^2+2\right)-2 \kappa ^2\right)-2 M n \left(n^2-3 n+2\right) r\right)+ {{f'}}^2 (n-2) n
   r^{n+2}\right)\right]^{-1}\nonumber\\
  c_0&=r^{-n-2} \left(2 f^2 \left(n^2-3 n+2\right) r^n \left(r^n \left(n^2 ( {{f'}} r-2)+n \left(-2  {{f'}}
  r+2 \kappa ^2+2\right)-2 \kappa ^2\right)-2 M n \left(n^2-3 n+2\right) r\right)\right.\nonumber\\
&\left.  +f \left(r^{2 n} \left(2 n^4
   \left( {{f'}}^2 r^2-3  {{f'}} r+2\right)+n^3 \left(-9  {{f'}}^2 r^2+2  {{f'}} \left(\kappa
   ^2+13\right) r-16\right)
+2 n^2 \left(6  {{f'}}^2 r^2- {{f'}} \left(5 \kappa ^2+18\right) r-2 \kappa
   ^4\right.
   \right.\right.\right.\nonumber\\
   &\left.\left.\left.\left.
    +4 \kappa ^2+10\right)  -4 n \left( {{f'}}^2 r^2-4  {{f'}} \left(\kappa ^2+1\right) r-2 \kappa ^4+4
      \kappa ^2+2\right)-4 \kappa ^2 \left(2  {{f'}} r+\kappa ^2-2\right)\right)
  \right.\right.\nonumber\\
   &\left.\left.
      +4 M \left(n^2-3 n+2\right)
   r^{n+1} \left(n^2 \left(3  {{f'}} r-\kappa ^2-9\right)+n \left(-4  {{f'}} r+5 \kappa ^2+6\right)-4
   \kappa ^2+3 n^3\right)-8 M^2 n \left(n^2-3 n+2\right)^2 r^2\right)+
\right.\nonumber\\
&\left.   {{f'}}^2 n r^{n+2} \left(r^n
   \left(n^2 (2  {{f'}} r+1)+n \left(-8  {{f'}} r-3 \kappa ^2+2\right)+4 \left(2  {{f'}} r+\kappa
   ^2-1\right)\right)+2 M \left(5 n^2-16 n+12\right) r\right)\right)\nonumber\\
  &\left[f \left(2 f (n-1) \left(r^n \left(n^2
   ( {{f'}} r-2)+n \left(-2  {{f'}} r+2 \kappa ^2+2\right)-2 \kappa ^2\right)-2 M n \left(n^2-3 n+2\right)
   r\right)+ {{f'}}^2 (n-2) n r^{n+2}\right)\right]^{-1}
\end{align}
\end{widetext}
and $\kappa^2=l(l+n-1)$.
%%%%%%%%%%%%%%%%%%%%%%%%%%%%%%%%%%%%%%%%%%%%%%%%%%%%%%%%%%%%%%%%%%%%%%%%%%%%%%

\bibliographystyle{h-physrev4}
\bibliography{References}

\end{document}